\documentclass[aps,twocolumn,showpacs,preprintnumbers]{revtex4}
\usepackage[nointlimits]{amsmath}
\usepackage{amsfonts}
\usepackage{amssymb}
\usepackage{amsmath}
\usepackage{amsthm}
\usepackage{latexsym}
\usepackage{graphicx}
\usepackage{color}
\usepackage{layout}
\usepackage[english]{babel}
\def\LL{{\cal L}}
\newcommand{\radiustar} {{\rm R}_{\rm S}}

\begin{document}

\title{Solar System constraints to nonminimally coupled gravity}

\author{Orfeu Bertolami\footnote{Also at Instituto de Plasmas e F\'isica Nuclear, Instituto Superior T\'ecnico, Av. Rovisco Pais, 1, 1049-001, Lisboa Portugal.}}
\email{orfeu.bertolami@fc.up.pt}
\affiliation{Departamento de F\'{\i}sica e Astronomia, Faculdade de Ci\^encias, Universidade do Porto,\\Rua do Campo Alegre 687,
4169-007 Porto, Portugal}

\author{Riccardo March}
\email{r.march@iac.cnr.it}
\affiliation{Istituto per le Applicazioni del Calcolo, CNR,\\Via dei Taurini 19, 00185 Roma, Italy,\\
and INFN - Laboratori Nazionali di Frascati (LNF),\\Via E. Fermi 40 Frascati, 00044 Roma, Italy}

\author{Jorge P\'aramos}
\email{paramos@ist.edu}
\homepage{web.ist.utl.pt/jorge.paramos}
\affiliation{Instituto de Plasmas e Fus\~ao Nuclear, Instituto Superior T\'ecnico\\Av. Rovisco Pais 1, 1049-001 Lisboa, Portugal}

\date{\today}

\begin{abstract}
We extend the analysis of Chiba, Smith and Erickcek \cite{CSE} of Solar System constraints on $f(R)$ gravity
to a class of nonminimally coupled (NMC) theories of gravity. These generalize $f(R)$ theories
by replacing the action functional of General Relativity (GR) with a more general form involving two
functions $f^1(R)$ and $f^2(R)$ of the Ricci scalar curvature $R$. While the function $f^1(R)$ is a nonlinear
term in the action, analogous to $f(R)$ gravity, the function $f^2(R)$ yields a NMC between
the matter Lagrangian density $\LL_m$ and the scalar curvature. The developed method allows for obtaining constraints on the admissible
classes of functions $f^1(R)$ and $f^2(R)$, by requiring that predictions of NMC gravity are compatible with
Solar System tests of gravity. We apply this method to a NMC model which accounts for the observed accelerated expansion of the Universe.
\end{abstract}


\maketitle

\section{Introduction}

One of the greatest challenges of contemporary physics is to make sense of the fact that, at Solar System level, there is no evidence that an extension of GR is required to account for all observed gravitational phenomena (see Ref. \cite{status} for a recent account), even though, from the theoretical point of view, GR is not a fully satisfactory theory. Indeed, GR exhibits singularities and is incompatible with Quantum Mechanics; furthermore, in order to account for the cosmological data, new states such as dark matter and dark energy are required.

As a possible alternative to this standard scenario, it is equally plausible that GR is actually an effective version of a more general theory of gravity. More recently, a great deal of interest has been dedicated to the so-called $f(R)$ theories \cite{DeT}; these can be further generalized by considering that matter and curvature are nonminimally coupled \cite{BBHL}, an idea that gives rise to many interesting features and has spanned several studies: these include the impact on stellar observables \cite{solarBP}, the so-called energy conditions \cite{BS}, the equivalence with multi-scalar-tensor theories \cite{equivalenceBP}, the possibility to account for galactic \cite{dm1BP} and cluster \cite{dm2BFP} dark matter, cosmological perturbations \cite{pertBFP}, a mechanism for mimicking a Cosmological Constant at astrophysical scales \cite{constantBP}, post-inflationary reheating \cite{reheating} or the current accelerated expansion of the universe \cite{BFP}, the dynamical impact of the choice of the Lagrangian density of matter \cite{BLP}, gravitational collapse \cite{collapse}, its Newtonian limit \cite{limitBM} and existence of closed timelike curves \cite{timelikeBF}.

In this work, we study whether a nonminimally coupled theory of gravity can be assessed using Solar System observables. It follows an analogous analysis, performed by Chiba, Smith and Erickcek \cite{CSE} for generic
$f(R)$ theories.
In Ref. \cite{CSE} the authors find a set of conditions that, when satisfied by the function $f(R)$,
lead to the prediction that the value of the parameterized post-Newtonian (PPN) parameter $\gamma$ is given by
$\gamma = 1 \slash 2$, which is not in agreement with Solar System tests of gravity.
Hence, the analysis of Ref. \cite{CSE} can be considered as a tool to rule out $f(R)$ theories that satisfy
a suitable set of conditions.
Particularly, it turns out that the $1 \slash R^n$ ($n>0$) gravity theory proposed by
Carroll-Duvvuri-Trodden-Turner \cite{CDTT} is ruled out by this analysis.

In the present paper we consider a class of NMC theories of gravity where
the action functional of GR is replaced with a more general form involving two
functions $f^1(R)$ and $f^2(R)$ of the Ricci scalar curvature $R$. The function $f^1(R)$ has a role
analogous to $f(R)$ gravity, and the function $f^2(R)$ yields a nonminimal coupling between
the matter Lagrangian density $\LL_m$ and the scalar curvature.

We extend the analysis of Ref. \cite{CSE} in order to develop a general framework for the study of Solar System constraints to NMC gravity. Then we apply the results of our analysis to a couple of case studies. Particularly, we consider the NMC model proposed by Bertolami, Fraz\~ ao and P\'aramos \cite{BFP} to account for the observed accelerated expansion of the Universe. This model posits an inverse power-law NMC $f^2(R) \propto 1 \slash R^n$ term in the action functional, and can be considered as a natural extension of $1 \slash R^n$ ($n>0$) gravity to a nonminimally coupled case. We show that, differently from pure $1 \slash R^n$ gravity, the NMC model of Ref. \cite{BFP} cannot be constrained or excluded by the method developed in this work. Hence such a NMC model remains, in this respect, a viable theory of gravity.

The manuscript is organized as follows: in sections II and III, we present our model and the assumptions adopted to ascertain the effect of the NMC in the Solar System. In sections IV and V, we carry out the suitable linearization of the relevant equations and derive the conditions required for applying the long range limit. Sections VI-VIII then address the solutions to the obtained set of equations. Section IX tackles the compatibility of the model under scrutiny with the various assumptions used to assess its impact at Solar System scales. Finally, we present our conclusions. An Appendix accounts for some technical aspects used to obtain the solution for linearized field equations.

\section{Nonminimally coupled gravity}

In the present work we consider gravitational theories with an action functional of the form \cite{BBHL},
\begin{equation}
S = \int \left[\frac{1}{2}f^1(R) + [1 +  f^2(R)] \LL_m \right]\sqrt{-g} \, d^4x,
\end{equation}
where $f^i(R)$ ($i=1,2$) are functions of the Ricci scalar curvature $R$, $\LL_m$ is the Lagrangian
density of matter and $g$ is the metric determinant. The standard Einstein-Hilbert action is recovered by taking
\begin{equation}
f^2(R) = 0, \qquad f^1(R) = 2\kappa (R - 2\Lambda),
\end{equation}
where $\kappa = c^4/16\pi G_N$ and $\Lambda$ is the Cosmological Constant. Here, $G_N$ is Newton's gravitational constant: as we will show, an effective gravitational constant $G$ arises due to the composite effect of $f^1(R)$ and $f^2(R)$.

The variation of the action functional with respect to the metric $g_{\mu\nu}$ yields the field equations
\begin{eqnarray}\label{field-eqs}
 && \left(f^1_R + 2 f^2_R \LL_m \right) R_{\mu\nu} - \frac{1}{2} f^1 g_{\mu\nu} =  \left(1 +  f^2 \right) T_{\mu\nu}\\ \nonumber&& +
\left(\nabla_\mu \nabla_\nu -g_{\mu\nu} \square \right) \left(f^1_R + 2 f^2_R \LL_m \right),
\end{eqnarray}
where $f^i_R \equiv df^i\slash dR$. In the following we assume that matter behaves as dust, {\it i.e.} a perfect fluid
with negligible pressure and an energy-momentum tensor described by
\begin{equation}
T_{\mu\nu} = \rho u_\mu u_\nu, \qquad u_\mu u^\mu = -1,
\end{equation}
where $\rho = \rho(r,t)$ is the matter density and $u_\mu$ is the four-velocity. The trace of the energy-momentum tensor
is $T = -\rho$. We use $\LL_m = -\rho$ for the Lagrangian density of matter (see Ref. \cite{BLP} for a discussion).

\section{Assumptions on the metric and on functions $f^1(R)$ and $f^2(R)$}
\label{sec:assumptions}

We now seek the metric that describes the spacetime around a spherical body such as the Sun in the weak-field limit of NMC gravity. Such a metric will be regarded as a perturbation of a background spacetime around which we linearize the field equations. We take the background metric to be a flat Friedmann-Robertson-Walker (FRW) metric
\begin{equation}
ds^2 = -dt^2 + a^2(t)(dr^2 + r^2 d\Omega^2),
\end{equation}
with scale factor $a(t)$ (we set $a(t)=1$ at the present time). Such a FRW metric solves the field Eqs. (\ref{field-eqs}) for a spatially uniform cosmological dust energy-momentum tensor, $T^{\rm cos}_{\mu\nu}$, the trace of which is $-\rho^{\rm cos}(t)$. We denote the Ricci scalar curvature of the background spacetime by $R_0 = R_0(t)$.

We assume that the spacetime around a spherical star is written (in spherical coordinates) by the following perturbation of the background metric,
\begin{eqnarray}\label{metric}
ds^2 &=& -\left[1 + 2\Psi(r,t) \right] dt^2 + \\ \nonumber && a^2(t)\left(\left[1 + 2\Phi(r,t)\right] dr^2
+ r^2 d\Omega^2 \right),
\end{eqnarray}
where $\Psi(r,t) \ll 1$ and $\Phi(r,t) \ll 1$. The Ricci curvature of the perturbed spacetime is expressed as the sum
\begin{equation}
R(r,t) = R_0(t) + R_1(r,t).
\end{equation}
As expected, we will show that the time scale of variations in $\Psi$, $\Phi$ and $R_1$ is much longer than the one
of Solar System dynamics, such that
\begin{equation}
\Psi(r,t) \simeq \Psi(r), ~~ \Phi(r,t) \simeq \Phi(r), ~~ R_1(r,t) \simeq R_1(r).
\end{equation}
Following Ref. \cite{CSE}, in the linearization of the field equations, both around and inside the star, we assume that
\begin{equation}\label{R1-cond}
R_1(r,t) \ll R_0(t).
\end{equation}
Such an assumption implies that the scalar curvature $R$ of the perturbed spacetime remains close to the cosmological value $R_0$ inside the star. In $f(R)$ theories this condition is satisfied, for instance, by the model proposed in Ref. \cite{CDTT}, where
\begin{equation}
f^1(R) = 2\kappa \left( R - \frac{\mu^4}{R}\right), \qquad f^2(R) = 0,
\end{equation}
as shown in Refs. \cite{CSE,HMV}. Such a behaviour for the curvature differs from the usual scenario of GR,
where the above condition breaks down inside the body, since the mass density of the star is larger than the cosmological mass density. This issue will play a central role in the application of the framework here developed to the NMC model proposed
in Ref. \cite{BFP}.
Naturally, the validity of condition Eq. (\ref{R1-cond}) will depend on the particular choice of $f^1(R)$ and $f^2(R)$, and thus can be used to constrain these functions.

We consider that all derivatives of functions $f^1(R)$ and $f^2(R)$ exist at the present value of $R_0(t)$. Since we assume that $R_1 \ll R_0$, we can Taylor expand $f^i(R)$ around $R=R_0$ to evaluate $f^i(R_0 + R_1)$ and $f^i_R(R_0 + R_1)$, for $i=1,2$. Neglecting terms nonlinear in $R_1$, we get
\begin{eqnarray}\label{condlinear}
f^i(R_0) + \frac{d f^i}{dR}(R_0) R_1 \gg \frac{1}{k!}\frac{d^k f^i}{dR^k}(R_0) R_1^k, \\ \nonumber
f^i_R(R_0) + \frac{d f^i_R}{dR}(R_0) R_1 \gg \frac{1}{k!}\frac{d^k f^i_R}{dR^k}(R_0) R_1^k,
\end{eqnarray}
for all $k>1$ and $i=1, 2$.
Following Ref. \cite{CSE}, we introduce the useful notation (for $i=1,2$),
\begin{equation}
f^i_0 \equiv f^i(R_0)~~,~~  f^i_{R0}  \equiv \frac{d f^i}{dR}(R_0)~~,~~ f^i_{RR0}  \equiv \frac{d^2 f^i}{dR^2}(R_0).
\end{equation}
%

\section{Linearization of the trace of the field equations}
\label{sec:linear-trace}

The trace of the field Eqs. (\ref{field-eqs}) is given by
\begin{eqnarray}\label{trace}
&& \left( f^1_R + 2 f^2_R \LL_m \right) R - 2f^1 + 3\square \left( f^1_R +2 f^2_R \LL_m \right) = \\ \nonumber && \left( 1 +  f^2 \right) T.
\end{eqnarray}
The energy-momentum tensor is decomposed in the following way:
\begin{equation}
T_{\mu\nu} = T_{\mu\nu}^{\rm cos} + T_{\mu\nu}^{\rm s}, \qquad \rho = \rho^{\rm cos} + \rho^{\rm s},
\end{equation}
where $\rho^{\rm cos}=\rho^{\rm cos}(t)$ is the cosmological matter density and $\rho^{\rm s}= \rho^{\rm s}(r)$ is the stellar matter density. The traces of the energy-momentum tensor contributions are denoted by $T^{\rm cos}$ and $T^{\rm s}$, respectively. We denote by $\radiustar$ the radius of the star
and assume that both the function $\rho^{\rm s}(r)$ and its derivative are continuous across the surface of the star, such that
\begin{equation}\label{star-boundary}
\rho^{\rm s}(\radiustar) = \frac{d\rho^{\rm s}}{dr}(\radiustar) = 0.
\end{equation}
We also write $\LL_m^{\rm cos} = -\rho^{\rm cos}$ and $\LL_m^{\rm s} = -\rho^{\rm s}$, so that $\LL_m = \LL_m^{\rm cos} + \LL_m^{\rm s}$.
As a consequence of our definitions, we have that $\rho(r,t) = \rho^{\rm cos}(t) + \rho^{\rm s}(r)$ inside the star.

The background curvature $R_0$ solves the trace Eq. (\ref{trace}) with matter source given by $T^{\rm cos}$:
\begin{eqnarray}\label{trace-R0}
\left( f^1_{R0} + 2 f^2_{R0} \LL_m^{\rm cos} \right) R_0 - 2f^1_0 + \\ \nonumber 3\square \left( f^1_{R0} +
2f^2_{R0} \LL_m^{\rm cos} \right) = \left( 1 +  f^2_0 \right) T^{\rm cos}.
\end{eqnarray}
We now linearize Eq. (\ref{trace}) using the first order Taylor expansions of the functions $f^i(R)$ and $f^i_R(R)$ around $R=R_0 \neq 0$. Since $R=R_0 + R_1$, using condition Eq. (\ref{R1-cond}), we neglect $O(R_1^2)$ contributions, but keep the cross-term $R_0 R_1$. Moreover, using the fact that $R_0$ solves Eq. (\ref{trace-R0}), we eliminate in the linearized trace equation terms that are independent of $R_1$, with the exception of those containing the matter source $T^{\rm s}=\LL_m^{\rm s}$. The application of the above procedure yields
\begin{eqnarray}\label{R1-equation}
&& \left[ -f^1_{R0} +  f^2_{R0}\LL_m + \left( f^1_{RR0} + 2 f^2_{RR0}\LL_m \right) R_0 \right] R_1 + \nonumber
 \\ && 3 \square \left[\left( f^1_{RR0} + 2 f^2_{RR0}\LL_m \right) R_1 \right] = \\ \nonumber
&& \left( 1 +  f^2_0 \right) T^{\rm s} - 2 f^2_{R0}\LL_m^{\rm s}R_0 -
6\square\left( f^2_{R0}\LL_m^{\rm s} \right).
\end{eqnarray}
In order to compute the term
\begin{equation}
\square \left[\left( f^1_{RR0} + 2 f^2_{RR0}\LL_m \right) R_1 \right],
\end{equation}
we consider the approximation $R_1(r,t) \simeq R_1(r)$, that will be verified later, obtaining
\begin{equation}
\square\left( f^1_{RR0} R_1\right) = f^1_{RR0} \square R_1 + R_1 \square f^1_{RR0},
\end{equation}
and
\begin{eqnarray}
&& \square \left( f^2_{RR0}\LL_m R_1 \right) = -f^2_{RR0}\rho^{\rm cos}\square R_1 - \\ \nonumber &&
R_1 \square \left( f^2_{RR0}\rho^{\rm cos} \right) -f^2_{RR0} \square \left (\rho^{\rm s}R_1 \right) -
\rho^{\rm s}R_1 \square f^2_{RR0}.
\end{eqnarray}
By definition,
\begin{eqnarray}
\square R_1(r) & = & g^{rr}\frac{d^2 R_1}{dr^2} - g^{\mu\nu}\Gamma^r_{~\mu\nu}\frac{dR_1}{dr},\\ \nonumber
\square \left( \rho^{\rm s}(r)R_1(r) \right) & = & g^{rr}\frac{d^2 \left( \rho^{\rm s}R_1 \right)}{dr^2} -
g^{\mu\nu}\Gamma^r_{~\mu\nu}\frac{d\left( \rho^{\rm s}R_1 \right)}{dr},
\end{eqnarray}
where $\Gamma^\lambda_{~\mu\nu}$ are the Christoffel symbols of the metric Eq. (\ref{metric}). Neglecting terms
in Eq. (\ref{R1-equation}) that
involve products of $R_1$ or its spatial derivatives with $\Psi$, $\Phi$ and their spatial derivatives (since such products turn out to be of order $o(1\slash c^2)$), we may approximate
\begin{equation}
\square R_1 \simeq \nabla^2 R_1, \qquad
\square\left( \rho^{\rm s}R_1\right) \simeq \nabla^2\left( \rho^{\rm s}R_1\right),
\end{equation}
where $\nabla^2$ denotes the three-dimensional flat space Laplacian. Taking into account that $f^2_{RR0} = f^2_{RR0}(t)$, it follows that
\begin{eqnarray}
&& \square \left( f^2_{RR0}\LL_m R_1 \right) \simeq \\ \nonumber &&
-R_1 \left[ \rho^{\rm s}\square f^2_{RR0} + \square \left( f^2_{RR0}\rho^{\rm cos} \right)\right] +
\nabla^2\left( f^2_{RR0}\LL_m R_1 \right).
\end{eqnarray}
Collecting these results, we thus find
\begin{eqnarray}
&& \square \left[\left( f^1_{RR0} + 2 f^2_{RR0}\LL_m \right) R_1 \right] \simeq \\ \nonumber && \left[ \square\left( f^1_{RR0} - 2 f^2_{RR0}\rho^{\rm cos}\right) -
2\rho^{\rm s}\square f^2_{RR0}\right] R_1 + \\ \nonumber && \nabla^2 \left[\left( f^1_{RR0} + 2 f^2_{RR0}\LL_m\right) R_1 \right].
\end{eqnarray}
The same steps are also applied to the term
\begin{equation}
\square\left( f^2_{R0}\LL_m^{\rm s} \right) = -f^2_{R0}\square\rho^{\rm s} - \rho^{\rm s}\square f^2_{R0},
\end{equation}
found in Eq. (\ref{R1-equation}); substituting the obtained expressions into Eq. (\ref{R1-equation}), we obtain
\begin{eqnarray}
&& 3\nabla^2 \left[\left( f^1_{RR0} + 2 f^2_{RR0}\LL_m\right) R_1 \right] + \\ \nonumber &&
\left( -f^1_{R0} +  f^2_{R0}\LL_m \right) R_1 + \left( f^1_{RR0} + 2 f^2_{RR0}\LL_m \right) R_0 R_1 + \\ \nonumber &&3 \left[ \square\left( f^1_{RR0} - 2 f^2_{RR0}\rho^{\rm cos}\right) - 2\rho^{\rm s}\square f^2_{RR0}\right] R_1 = \\ \nonumber && -\left( 1 +  f^2_0 \right) \rho^{\rm s} + 2 f^2_{R0}\rho^{\rm s}R_0 +
6 \rho^{\rm s} \square f^2_{R0} + 6 f^2_{R0} \nabla^2 \rho^{\rm s}.
\end{eqnarray}
We define the potential
\begin{equation}\label{potentialdef}
U(r,t) = \left[ f^1_{RR0}(t) + 2 f^2_{RR0}(t)\LL_m(r,t)\right] R_1(r),
\end{equation}
and the mass parameter
\begin{eqnarray}\label{mass-formula}
m^2 &=&  \frac{1}{3}\bigg[\frac{f^1_{R0} -  f^2_{R0}\LL_m}{f^1_{RR0} + 2 f^2_{RR0}\LL_m}
- R_0 - \\ \nonumber &&  \frac{3\square\left( f^1_{RR0} - 2 f^2_{RR0}\rho^{\rm cos}\right)
- 6\rho^{\rm s}\square f^2_{RR0}}{f^1_{RR0} + 2 f^2_{RR0}\LL_m} \bigg],
\end{eqnarray}
 assuming that $f^1_{RR0} + 2 f^2_{RR0}\LL_m \neq 0$. Note that $m = m(t)$ outside the spherical body, where $\rho^s = 0$ and $\LL_m = - \rho^{cos}(t)$. For $f^2(R)=0$, the mass formula presented in Ref. \cite{CSE} for $f(R)$ theories is recovered. A negative mass squared $m^2 < 0$ could generically produce a gravitational instability, as the solution of Eq. (\ref{U-equation-simplified}) would lead to radial oscillations of the potential $U$ with wavelength and frequency $\sim |m|^{-1}$.

In the remainder of this study, we will assume that $|mr| \ll 1$ within the Solar System, so that the contribution of any mass parameter is negligible and any putative oscillations evolve with a wavelength and period much larger than the typical timescale of Solar System dynamics.

Using the expressions for $U$ and $m^2$, the equation for $R_1$ can be written as
\begin{eqnarray}\label{U-equation}
&& \nabla^2 U -m^2 U = \\ \nonumber &&-\frac{1}{3}\left( 1 +  f^2_0 \right) \rho^{\rm s} +
\frac{2}{3} f^2_{R0}\rho^{\rm s}R_0 + 2 \rho^{\rm s} \square f^2_{R0} +
2 f^2_{R0} \nabla^2 \rho^{\rm s}.
\end{eqnarray}

 The assumption $|m r| \ll 1$ at Solar System scales signals a long-range extra force due to the non-trivial functions $f^i(R)$. If the mass parameter is negative, this implies that the timescale of oscillations is much larger than the one ruling Solar System dynamics.

\section{Solution for $R_1$}

Outside the star, Eq. (\ref{U-equation}) reads $\rho^{\rm s} = 0$ and we obtain
\begin{equation} \nabla^2 U = m^2(t) U,\end{equation}
so that $U$ behaves as a Yukawa potential with a characteristic length $1/m(t)$ evolving on a cosmological timescale,
\begin{equation}
U \sim \frac{ e^{-mr}}{r} \sim \frac{1}{r},
\end{equation}
or, if $m^2$ is negative, as an oscillating potential with strength $\sim 1/r$. The approximation $U\sim 1/r$ stems from the assumption that $|mr| \ll 1$ within the Solar System: we may thus drop the mass term $m^2 U$ in Eq.  (\ref{U-equation}) outside the spherical body.
Moreover, standard approximation properties of solutions of differential equations permit us to neglect this mass
term also inside the spherical body, where the mass $m^2$ depends both on $r$ and $t$, whenever $|mr| \ll 1$.
Eq. (\ref{U-equation}) then becomes
\begin{equation}\label{U-equation-simplified}
\nabla^2 U = \eta(t)\rho^{\rm s}(r) + 2 f^2_{R0} \nabla^2 \rho^{\rm s},
\end{equation}
with the definition
\begin{equation}\label{etadef}
\eta(t) = -\frac{1}{3}\left( 1 + f^2_0 \right) +
\frac{2}{3} f^2_{R0}R_0 + 2 \square f^2_{R0}.
\end{equation}
Outside the spherical body, $\rho^{\rm s}=0$ and we may use the divergence theorem to obtain
\begin{equation}\label{U-equation-simplified-outside}
U(r,t) = -\frac{\eta(t) }{ 4\pi} \frac{M_{\rm S}}{ r},
\end{equation}
where $M_{\rm S}$ is the total gravitational mass of the spherical body. Using Eq. (\ref{potentialdef}), this implies that
\begin{equation}\label{R1-solution}
R_1(r,t) = \frac{\eta(t) }{ 4\pi \left( 2 f^2_{RR0}\rho^{\rm cos}-  f^1_{RR0}\right)} \frac{M_{\rm S}}{ r}.
\end{equation}
For $f^2(R)=0$, this expression reduces to the solution for $R_1$ found in Ref. \cite{CSE}.
Notice that, although $R_1$ depends on time through $R_0(t)$ and $\rho^{\rm cos}(t)$, the timescale of its variation (comparable to the current Hubble time being much bigger than the one of Solar System dynamics) ensures the approximation $R_1(r,t) \simeq R_1(r)$.

Inside the spherical body, Eq. (\ref{U-equation-simplified}) implies that
\begin{equation}\label{U-equation-simplified-inside}
\frac{d }{ dr}\left(U - 2 f^2_{R0} \rho^{\rm s} \right) = \frac{\eta(t) }{ 4\pi} \frac{M(r) }{ r^2},
\end{equation}
where $M(r)$ is the gravitational mass inside a sphere of radius $r$, defined as
\begin{equation}
M(r) \equiv 4\pi \int_0^r \rho^{\rm s}(\xi) \xi^2 d\xi~~,~~M_{\rm S} = M(\radiustar).
\end{equation}
Since the potential $U$ must be continuous, it is profitable to rewrite this equation in terms of the dimensionless variable $x \equiv r/\radiustar$ and dimensionless function
\begin{equation}\label{y-definition}
y\equiv \frac{U(x)}{U(x=1)} = -\frac{4\pi \radiustar U(x)}{\eta(t)M_{\rm S} },
\end{equation}
so that Eq. (\ref{U-equation-simplified-inside}) becomes
\begin{equation}\label{U-equation-simplified-inside-dim}
\frac{d }{ dx}\left( y + \frac{8\pi f^2_{R0} }{\eta(t) }\frac{\radiustar}{M_{\rm S}} \rho^{\rm s} \right) = -\frac{M(x) }{ M_{\rm S} x^2 }.
\end{equation}
In order to derive $y(x)$ from the above, we require prior knowledge of the density profile inside the spherical body, $\rho^{\rm s}$; to do so, we assume that the latter may be expanded as a Taylor series,
\begin{equation}
\label{rhoTaylor} \rho^{\rm s } = \rho^{\rm s}_0 \sum_{i=0} a_i x^i ,
\end{equation}
where $\rho^{\rm s}_0 \sim 10^5~{\rm kg/m^3}$ is the central density and $a_0 = 1$. We thus get
\begin{equation} M (r) = 4\pi \rho^{\rm s}_0 \radiustar^3 \sum_{i=0} \frac{a_i}{i+3}x^{i+3} ,
\end{equation}
so that
\begin{equation}\label{total-mass}
M_{\rm S} = 4\pi \rho^{\rm s}_0 \radiustar^3 \sum_{i=0} \frac{a_i}{i+3} ,
\end{equation}
and Eq. (\ref{U-equation-simplified-inside-dim}) may be integrated between $x$ and $x=1$ to obtain
\begin{equation}\label{U-equation-simplified-inside-dim-solution}
y = \frac{ \sum_{i=0} \frac{a_i}{i+2} }{ \sum_{i=0} \frac{a_i}{i+3} } - \frac{\sum_{i=0} a_i x^i \left[ \frac{2 f^2_{R0} }{\eta(t) \radiustar^2} + \frac{x^2}{(i+2)(i+3)} \right] }{ \sum_{i=0} \frac{a_i}{i+3}}.
\end{equation}
Using Eqs. (\ref{U-equation-simplified-inside-dim-solution}) and (\ref{potentialdef}), we thus obtain
\begin{equation}\label{perturbativecondition}
\frac{R_1}{R_0} = \frac{\eta }{ 4\pi [ 2 f^2_{RR0} (\rho^{\rm cos} + \rho^{\rm s}) - f^1_{RR0} ]} \frac{M_{\rm S} }{R_0 \radiustar}y.
\end{equation}
Eq. (\ref{perturbativecondition}) must be used to check if the perturbative approach $R_1 \ll R_0$ is valid within the spherical body. Outside it, it suffices to compare Eq. (\ref{R1-solution}) with the expression for $R_0$ found from a cosmological solution of NMC gravity.

The condition $R_1 \ll R_0$ implies that the Ricci curvature $R=R_0 + R_1$ of the perturbed spacetime
is close to the cosmological value $R_0$ at Solar System scales, and also inside the spherical body,
even though the metric Eq. (\ref{metric}) of the perturbed spacetime is fairly close to the Minkowski metric.

In theories where $f^2(R)=0$, such a condition is satisfied for $mr \ll 1$, with $r$ varying from Solar System scales to the star interior,
and $f^1_{R0}\slash f^1_{RR0} \sim R_0$ \cite{CSE,DeT}. However, such theories yield the value $\gamma = 1\slash 2$
which does not satisfy Solar System tests of gravity. Theories which do not satisfy the condition
$R_1 \ll R_0$ inside the spherical body are characterized by a large mass $m$, such that $mr \gg 1$ at Solar System scales \cite{DeT}.
For $f^2(R)=0$, this could render viable, due to decoupling, a minimally coupled model of gravity; for GR, the condition $R_1 \ll R_0$ is not satisfied in the star interior.
In this study, we consider this issue for $f^2(R) \neq 0$.

\section{Linearization of the field equations}

In this section we linearize the field Eqs. (\ref{field-eqs}). We denote by $\left[ R_0 \right]^\mu_\nu$
the components of the Ricci tensor in the considered background metric.
The tensor $\left[ R_0 \right]^\mu_\nu$ solves the field Eqs. (\ref{field-eqs}) with matter source given
by $T^{{\rm cos}\,\mu}_\nu$:
\begin{eqnarray}\label{backg-eqs}
&& \left( \left[ R_0 \right]^\mu_\nu -\nabla^\mu \nabla_\nu + \delta^\mu_\nu \square \right)
\left( f^1_{R0} + 2 f^2_{R0}\LL_m^{\rm cos} \right) - \\ \nonumber && \frac{1}{2}f^1_0 \delta^\mu_\nu
= \left( 1 +  f^2_0 \right)T^{{\rm cos}\,\mu}_\nu.
\end{eqnarray}
We now linearize Eqs. (\ref{field-eqs}) using the first order Taylor expansions of the functions $f^i(R)$ and $f^i_R(R)$
around $R=R_0$, for $i=1,2$. Using Eq. (\ref{backg-eqs}) and neglecting time derivatives of the background metric, we obtain the following system of equations in $R^\mu_\nu$:
\begin{eqnarray}\label{Rtensor-eqs}
& &\left( f^1_{R0} + 2 f^2_{R0}\LL_m^{\rm cos} \right)
\left( R^\mu_\nu - \left[ R_0 \right]^\mu_\nu \right) + 2 f^2_{R0}\LL_m^{\rm s}R^\mu_\nu + \\ \nonumber &&
\left( f^1_{RR0} + 2 f^2_{RR0}\LL_m \right) R_1R^\mu_\nu -  f^2_{R0}R_1 T^\mu_\nu - \frac{1}{2}f^1_{R0} R_1\delta^\mu_\nu - \\ \nonumber &&
f^1_{RR0}\left( \nabla^\mu \nabla_\nu -\delta^\mu_\nu \square \right) R_1
- 2 f^2_{RR0}\left( \nabla^\mu\nabla_\nu - \delta^\mu_\nu \square \right)\left( \LL_m R_1 \right) \\ \nonumber
&&= \left( 1 +  f^2_0 \right)T^{{\rm s}\,\mu}_\nu + 2 f^2_{R0}\left( \nabla^\mu\nabla_\nu - \delta^\mu_\nu \square \right) \LL_m^{\rm s}.
\end{eqnarray}
The $R^0_0$ component is thus given by
\begin{eqnarray}
R^0_0 &=& -\frac{1}{1 + 2\Psi}\left[ \frac{1}{a^2}\nabla^2\Psi - 3\left( H^2 + \frac{dH}{dt}\right) \right] \simeq \\ \nonumber &&
- \nabla^2\Psi + 3\left( H^2 + \frac{dH}{dt}\right),
\end{eqnarray}
while $R_{rr}$ reads
\begin{eqnarray}
R_{rr} &=& a^2 \frac{1 + 2\Phi}{1 + 2\Psi}\left( 3H^2 + \frac{dH}{dt}\right) - \frac{1}{1 + 2\Psi}\frac{d^2\Psi}{dr^2} + \\ \nonumber && \frac{1}{\left( 1 + 2\Psi\right)^2}\left( \frac{d\Psi}{dr} \right)^2 + \frac{2}{r}\frac{1}{1 + 2\Phi}\frac{d\Phi}{dr} + \\ \nonumber && \frac{1}{\left(1+2\Phi\right)\left(1+2\Psi\right)}\frac{d\Phi}{dr}\frac{d\Psi}{dr} \\ \nonumber
&\simeq & - \frac{d^2\Psi}{dr^2} + \frac{2}{r}\frac{d\Phi}{dr} + 3H^2 + \frac{dH}{dt}.
\end{eqnarray}
By neglecting the terms involving functions $\Psi$ and $\Phi$ in the previous expressions we get the corresponding
components of the tensor $\left[ R_0 \right]^\mu_\nu$.

We can simplify Eqs. (\ref{Rtensor-eqs}) by neglecting terms involving the product of $R_1$, $\Psi$, $\Phi$ and their
derivatives with $H$ and $dH\slash dt$. Moreover, following Ref. \cite{CSE}, we neglect terms that are nonlinear
functions of the metric perturbations $\Psi$ and $\Phi$, and we neglect terms involving products of $R_1$ by $\Psi$ and $\Phi$. Such approximations permit us to replace the D'Alembert operator $\square$ with the flat space Laplace
operator $\nabla^2$. The $00$ and $rr$ components of Eqs. (\ref{Rtensor-eqs}) then become, respectively,
\begin{eqnarray}\label{Psi-equation}
&& \left( f^1_{R0} + 2 f^2_{R0}\LL_m \right) \left( \nabla^2\Psi + \frac{1}{2}R_1 \right) - \\ \nonumber && \nabla^2 \left[ \left( f^1_{RR0} + 2 f^2_{RR0}\LL_m \right) R_1 \right] \\ \nonumber &=& \left( 1 +  f^2_0 \right) \rho^{\rm s} - 2 f^2_{R0} \nabla^2 \rho^{\rm s},
\end{eqnarray}
and
\begin{eqnarray}\label{Phi-equation}
&& \left( f^1_{R0} + 2 f^2_{R0}\LL_m \right)\left( -\frac{d^2\Psi}{dr^2} + \frac{2}{r}\frac{d\Phi}{dr} \right) - \\ \nonumber && \frac{1}{2}f^1_{R0}R_1 + \frac{2}{r}f^1_{RR0}\frac{dR_1}{dr} + \frac{4}{r} f^2_{RR0}\frac{\partial\left(\LL_m R_1\right)}{\partial r} \\ \nonumber
&=& \frac{4}{r} f^2_{R0}\frac{d\rho^{\rm s}}{dr}.
\end{eqnarray}
In the next sections we shall compute the solutions $\Psi$ and $\Phi$ of these equations.

\section{Solution for $\Psi$}
\label{sec:sol-Psi}

Using Eqs. (\ref{U-equation-simplified}) and (\ref{etadef}), equation (\ref{Psi-equation}) becomes
\begin{eqnarray}
&& \left( f^1_{R0} + 2 f^2_{R0}\LL_m \right) \left( \nabla^2\Psi + \frac{1}{2}R_1 \right) = \\ \nonumber &&
\frac{2}{3} \left(  1 +  f^2_0 +  f^2_{R0}R_0 \right) \rho^{\rm s},
\end{eqnarray}
where we have neglected the term $\square f^2_{R0}$ on timescales of Solar System dynamics.

We assume that $f^1_{R0} + 2 f^2_{R0}\LL_m \neq 0$ and, following Ref. \cite{CSE}, decompose $\Psi$ as the sum of two functions, $\Psi = \Psi_0 + \Psi_1$, such that
\begin{eqnarray}\label{eqpsi}
\nabla^2 \Psi_0 &=& \frac{2}{3} \,\frac{1 +  f^2_0 +  f^2_{R0}R_0}
{f^1_{R0} + 2 f^2_{R0}\LL_m} \,\rho^{\rm s}, \\ \nonumber
\nabla^2 \Psi_1 &=& - \frac{1}{2} R_1.
\end{eqnarray}
Using Eq. (\ref{star-boundary}), integration through the divergence theorem yields for the function $\Psi_0$ outside of the star,
\begin{eqnarray}
\Psi_0(r,t) &=& -\frac{1}{6\pi}\left( 1 +  f^2_0 +  f^2_{R0}R_0 \right)
\frac{M^\ast}{r} + C_0, \\ \nonumber
M^\ast &=& 4\pi \int_0^{\radiustar} \frac{\rho^{\rm s}(x)}{f^1_{R0} + 2 f^2_{R0}\LL_m(x)}r^2\, dr,
\end{eqnarray}
with $C_0$ an integration constant. The function $\Psi_1$ is computed in the Appendix, where it is shown that, under the additional condition
\begin{equation}\label{constraint}
\left\vert\frac{f^1_{R0} + 2 f^2_{R0}\LL_m}{f^1_{RR0} + 2 f^2_{RR0}\LL_m}\right\vert
\sim \left\vert R_0 \right\vert,
\end{equation}
assumed to be valid both inside and outside the star, we have
\begin{eqnarray}
\Psi_1(r,t) &=& \Psi_1^\ast(r,t) + C_1, \\ \nonumber
\left\vert \Psi_1^\ast(r,t) \right\vert &\ll& \left\vert \Psi_0(r,t) - C_0 \right\vert,
\end{eqnarray}
where $C_1$ is another integration constant. Condition Eq. (\ref{constraint}) is satisfied for instance by
functions of the type $f^1(R) \sim R^m$, $f^2(R) \sim R^n$,
and its meaning will be discussed at the end of this section.
By requiring that $\Psi(r,t)$ vanishes as $r \rightarrow +\infty$, we obtain that $C_0 + C_1 = 0$.
The validity of the Newtonian limit requires that $\Psi(r)$ is proportional to $M \slash r$, leading to the following constraint on the functions $f^1(R)$ and $f^2(R)$:
\begin{equation}\label{Newt-limit}
\left\vert 2 f^2_{R0} \right\vert \rho^{\rm s}(r) \ll
\left\vert f^1_{R0} - 2 f^2_{R0}\rho^{\rm cos}(t) \right\vert, \qquad r \leq \radiustar.
\end{equation}
We now get the solution for $\Psi$ outside of the star,
\begin{equation}\label{Psi-solution}
\Psi(r,t) = - \frac{ 1 +  f^2_0 +  f^2_{R0}R_0 }
{6\pi \left( f^1_{R0} - 2 f^2_{R0}\rho^{\rm cos} \right)} \,\frac{M}{r},
\qquad r \geq \radiustar.
\end{equation}
For $f^2(R)=0$, this expression reduces to the solution for $\Psi$ found in Ref. \cite{CSE}.
The expression for $\Psi$ yields a gravitational coupling slowly varying in time,
\begin{eqnarray}\label{G-constant}
G &=& \frac{ \omega(t) }
{6\pi \left( f^1_{R0} - 2 f^2_{R0}\rho^{\rm cos} \right)}, \\ \nonumber
\omega(t) &=& 1 +  f^2_0 +  f^2_{R0}R_0.
\end{eqnarray}
As expected, the timescale $\dot{G}/G$ is much longer than the one of Solar System dynamics. Hence we have approximately $G \simeq {\rm const.}$ and $\Psi(r,t) \simeq \Psi(r)$.

By comparing with available bounds on $\dot{G}/G$  (see Ref. \cite{ChibaReview} for an updated review), Eq. (\ref{G-constant}) can in principle be used to constraint $f^1(R)$ and $f^2(R)$.

We may now check the assumption $R_1 \ll R_0$ outside the spherical body.
Using the solution Eq. (\ref{R1-solution}) for $R_1$ and the expression Eq. (\ref{G-constant}) of the effective gravitational constant $G$, we have, for $r \geq \radiustar$,
\begin{equation}
\left\vert\frac{R_1}{R_0}\right\vert \leq \,
\frac{3}{2\left\vert R_0 \right\vert}\frac{GM_{\rm S}}{\radiustar}
\left\vert\frac{\eta(t)}{\omega(t)}\right\vert \cdot
\left\vert\frac{f^1_{R0} - 2 f^2_{R0}\rho^{\rm cos}}{f^1_{RR0} - 2 f^2_{RR0}\rho^{\rm cos}}\right\vert.
\end{equation}
Then, the assumption $R_1 \ll R_0$ used in the linearization of the field equations places the following additional
constraint on functions $f^1(R)$ and $f^2(R)$,
\begin{equation}\label{add-constraint}
\left\vert\frac{\eta(t)}{\omega(t)}\right\vert \cdot
\left\vert\frac{f^1_{R0} - 2 f^2_{R0}\rho^{\rm cos}}{f^1_{RR0} - 2 f^2_{RR0}\rho^{\rm cos}}\right\vert
\ll \left\vert R_0 \right\vert \left(\frac{\radiustar}{GM_{\rm S}}\right).
\end{equation}
Neglecting the term $\square f^2_{R0}$ in $\eta(t)$ at the time-scale of Solar System dynamics, if
$\left\vert \eta(t) \slash \omega(t) \right\vert \sim 1$, then condition Eq. (\ref{constraint}) becomes
a sufficient condition for the validity of the assumption $R_1 \ll R_0$ outside the spherical body.
Indeed, if condition Eq. (\ref{constraint}) is satisfied and the effective gravitational constant $G$
is identified with Newton's gravitational constant, using Eq. (\ref{add-constraint}) we have
$\left\vert R_1 \slash R_0\right\vert \lesssim GM_{\rm S} \slash \radiustar \ll 1$.
For $f^2(R)=0$, condition Eq. (\ref{constraint}) reduces to condition $f^1_{R0}\slash f^1_{RR0} \sim R_0$
found in Ref. \cite{CSE}. This condition is satisfied for instance by the theory of $1\slash R^n$ gravity, proposed in Ref. \cite{CDTT}, where
\begin{eqnarray}\label{carroll}
f^1(R) &=& 2\kappa \left( R - \frac{\mu^{2+2n}}{R^n}\right), \quad n>0, \\ \nonumber f^2(R) &=& 0.
\end{eqnarray}
This theory satisfies also the condition $mr \ll 1$ at Solar System scales \cite{CSE}.

\section{Solution for $\Phi$}
\label{sec:sol-Phi}

We now compute the solution $\Phi$ under condition Eq. (\ref{constraint}). For $r \geq \radiustar$, Eq. (\ref{Phi-equation}) becomes
\begin{eqnarray}\label{Phi-eq-outside}
&& \left( f^1_{R0} + 2 f^2_{R0}\LL_m^{\rm cos} \right)\left( -\frac{d^2\Psi}{dr^2}
+ \frac{2}{r}\frac{d\Phi}{dr} \right) - \\ \nonumber && \frac{1}{2}f^1_{R0}R_1 + \frac{2}{r}\left(
f^1_{RR0} + 2 f^2_{RR0}\LL_m^{\rm cos} \right) \frac{dR_1}{dr} = 0.
\end{eqnarray}
Using the solution Eq. (\ref{R1-solution}) for $R_1$, we have
\begin{equation}
\frac{R_1}{dR_1\slash dr} = -r.
\end{equation}
Since $\rho^{\rm cos}(t) \ll \rho^{\rm s}(r)$ for $r < \radiustar$ and $\left\vert r - \radiustar \right\vert$ large enough, using Eq. (\ref{Newt-limit}) we have also
\begin{equation}\label{rhocos-small}
\left\vert 2 f^2_{R0} \right\vert \rho^{\rm cos}(t) \ll \left\vert f^1_{R0} \right\vert.
\end{equation}
Using these results and Eq. (\ref{constraint}), we have
\begin{eqnarray}
&& \left\vert \frac{f^1_{R0}R_1 \slash 2}{(2\slash r)
\left( f^1_{RR0} + 2 f^2_{RR0}\LL_m^{\rm cos} \right)\left(dR_1\slash dr\right)} \right\vert \simeq \\ \nonumber &&
\frac{1}{4}\left\vert \frac{f^1_{R0} + 2 f^2_{R0}\LL_m^{\rm cos}}
{f^1_{RR0} + 2 f^2_{RR0}\LL_m^{\rm cos}} \right\vert \cdot \left\vert \frac{R_1}
{(1\slash r)\left(dR_1\slash dr\right)} \right\vert \ll 1,
\end{eqnarray}
where we have used $\left\vert R_0 \right\vert r^2 \sim H^2 r^2 \ll 1$, for the current Hubble parameter $H$ and for $r$ of the order of Solar System scales.
It follows that the term $f^1_{R0}R_1 \slash 2$ can be neglected in Eq. (\ref{Phi-eq-outside}), which now becomes
\begin{equation}
\frac{d\Phi}{dr} = \frac{r}{2}\, \frac{d^2\Psi}{dr^2} - \left[\frac{f^1_{RR0} - 2 f^2_{RR0}\rho^{\rm cos}}
{f^1_{R0} - 2 f^2_{R0}\rho^{\rm cos}}\right] \, \frac{dR_1}{dr}.
\end{equation}
Substituting in this equation the derivatives of functions $R_1$ and $\Psi$, computed from
Eqs. (\ref{R1-solution}) and (\ref{Psi-solution}), respectively, we obtain
\begin{equation}
\Phi(r,t) = \frac{1 +  f^2_0 + 4 f^2_{R0}R_0 + 6\square f^2_{R0}}
{12\pi \left( f^1_{R0} - 2 f^2_{R0}\rho^{\rm cos} \right)} \,\frac{M}{r},
\end{equation}
for $ r \geq \radiustar$. As expected, setting $f^2(R)=0$ reduces this expression to the solution for $\Phi$ found in Ref. \cite{CSE}.
Again, we have $\Phi(r,t) \simeq \Phi(r)$.

Using the expressions of $\Psi$ and $\Phi$, we get the PPN parameter $\gamma$:
\begin{equation}\label{gamma}
\gamma = \frac{1}{2} \, \left[\frac{1 +  f^2_0 + 4 f^2_{R0}R_0 + 6\square f^2_{R0}}
{1 +  f^2_0 +  f^2_{R0}R_0} \right].
\end{equation}
Thus, the parameter $\gamma$ is completely defined by the background metric and its value can be obtained by computing the cosmological solution of NMC gravity. Inserting $f^2(R)=0$ yields the value $\gamma = 1\slash 2$ as it has been found in Ref. \cite{CSE}. In particular, the $1\slash R^n$ gravity model given by Eq. (\ref{carroll}) also predicts
$\gamma = 1\slash 2$.
However, notice that formula (\ref{gamma}) cannot be applied when the functions $f^i (R)$ reduce to their GR expressions, since in this case the mass parameter $m$, defined in Eq. (\ref{mass-formula}), is ill-defined (and divergent), so that the assumptions of our computations are not satisfied.

We may now summarize the obtained results: in order for a cosmologically viable nonminimally coupled model to be compatible with Solar System tests, one of the following conditions has to be satisfied:
\begin{itemize}
\item[{\rm (i)}] Either the condition $|mr| \ll 1$ at Solar System scales is not satisfied, or nonlinear terms in $R_1$ are not negligible in the Taylor expansions Eqs. (\ref{condlinear}) (which happens if the perturbative condition $R_1 \ll R_0$ is not satisfied), so that the present analysis does not apply;
\item[{\rm (ii)}] If both conditions of point (i) are satisfied, then the condition Eq. (\ref{Newt-limit}) of validity of the Newtonian limit has to be satisfied, and the value of $\gamma$ given by Eq. (\ref{gamma}) has to satisfy the constraint from the Cassini measurement $\gamma = 1 + (2.1 \pm 2.3) \times 10^{-5}$ ({\it cf.} Ref. \cite{status}).
\end{itemize}
The mass $m^2$, which is a function $m^2 = m^2(r,t)$ given by Eq. (\ref{mass-formula}), has to be computed by using the cosmological solution $R_0(t), \rho^{\rm cos}(t)$. In the following section, we implement the obtained criteria for the cosmological scenario posited in Ref. \cite{BFP}.

\section{Application}
\label{compatibility}

Following Ref. \cite{BFP}, let us consider the case study
\begin{equation}\label{case-study}
f^1(R) = 2\kappa R, \qquad f^2(R) = \left( \frac{R}{R_n} \right)^{-n}, \quad n>0,
\end{equation}
where $R_n$ is a constant; the linear choice of $f^1(R)$ serves to highlight the impact of the NMC between matter and curvature on the dynamics. Notice that the correct GR limit of a power-law coupling between matter and curvature is not attained by setting $n=0$ (as this simply doubles the minimal coupling, $f^2(R) = 1$), but by imposing $R_n \rightarrow 0$ (for positive $n$, {\it i.e.} an inverse power-law).

The above choice yields a cosmological scenario where the contribution of the NMC dominates the dynamics and a constant (negative) deceleration parameter is obtained, $q <0$; this, however, is attained due to the large value of $f^2_{R0}\rho^{\rm cos}$ and its temporal derivatives, not the NMC itself, which remains subdominant, $f^2_0 \ll 1$.

This mechanism implies a direct relation between the exponent $n$ and the latter \cite{BFP},
\begin{equation}
q = -1 + \frac{3 }{ 2(1+n)},
\end{equation}
that is, a De Sitter solution with exponential scale factor is ruled out. Thus, the scale factor $a(t)$ of the background metric and the cosmological matter density $\rho^{\rm cos}(t)$ follow the temporal evolution
\begin{equation}\label{behaviourBFP}
a(t) = a_0 \left( \frac{t}{t_0} \right)^{2(1+n)/3},
\end{equation}
and
\begin{eqnarray}\label{rhocosmo}
\rho^{\rm cos} (t) &=& \rho^{\rm cos}_0 \left( \frac{t_0}{t}\right)^{2(1+n)}, \\ \nonumber \rho_0 &=& (1+n)\frac{8}{3} \frac{\kappa}{t_0^2}\left(\frac{4(1+n)(1+4n)}{3R_n t_0^2}\right)^n ,
\end{eqnarray}
where $t_0$ is the current age of the Universe; the latter expression stems from the covariant conservation of the energy-momentum tensor, which remains valid since the Lagrangian density is given by $\LL_m = -\rho^{\rm cos}$ (see Ref. \cite{BLP} for a discussion).

Eq. (\ref{behaviourBFP}) yields
\begin{eqnarray}\label{constantdeceleration}
H &=& \frac{\dot{a}}{a} = \frac{2(1+n)}{3t}, \\ \nonumber  R_0 &=& 6( \dot{H} + 2H^2) = \frac{4(1+4n) (1+n) }{3t^2} ,
\end{eqnarray}
where $H \equiv \dot{a}/a$ is the Hubble parameter. Since the current value of the former is $H_0 \sim 70 ~{\rm (Km/s)/Mpc}$ \cite{WMAP9} and the deceleration parameter is of order $q_0 \sim -1 $, we get that $R_0 \sim (10^{14}~{AU})^{-2}$ --- to be compared with the relevant range for the Solar System, $r \lesssim {\rm R}_{SS} \sim 100~{\rm ~AU}$.

Inserting the expression for the scalar curvature $R_0$ into Eq. (\ref{case-study}), we get
\begin{equation}
f^2_0 = \left[ \frac{3}{4(4n+1)(n+1)} \left(\frac{t_0}{t_n}\right)^2 \right]^n,
\end{equation}
where $t_n \equiv 1/\sqrt{R_n}$.

We recall that the choice for the Lagrangian density $\LL_m = -\rho^{\rm cos}$ implies that the energy-momentum tensor of matter is conserved, $\nabla_\mu T^{\mu\nu} = 0 \rightarrow \dot{\rho}^{\rm cos} = -3 H \rho^{\rm cos}$. From Eq. (\ref{constantdeceleration}), we get
\begin{eqnarray}\label{R0behaviour}
\dot{R}_0 &=& - \frac{2}{t}R_0~~~~,~~~~\ddot{R}_0 =  \frac{6}{t^2}R_0,
\end{eqnarray}
and, together with the expressions below, valid for a power-law NMC,
\begin{equation}\label{expressionsf2}f^2_{R0} = -n \frac{f^2_0}{R_0}~~~~,~~~~f^2_{RR0} = n(n+1)\frac{f^2_0}{R_0^2} ,\end{equation}
we get
\begin{eqnarray}\label{dalemb-1}
&& \frac{\square\left( f^2_{RR0}\rho^{\rm cos}\right)}{ f^2_{RR0}\rho^{\rm cos}} \approx \\ \nonumber && -\frac{R_0^2}{ f^2_0\rho^{\rm cos}} \left[ \frac{d^2}{dt^2} \left(\frac{f^2_0}{R_0^2}\rho^{\rm cos}\right) + 3H\frac{d}{dt} \left(\frac{f^2_0}{R_0^2}\rho^{\rm cos}\right) \right] = \\ \nonumber && -\frac{3}{2}\frac{2n+3}{4n^2+5n+1} R_0 ,\end{eqnarray}
and
\begin{eqnarray}\label{dalemb-2}
\frac{\square f^2_{RR0}}{f^2_{RR0}} &\approx&  - \frac{R_0^2}{f^2_0}\left[ \frac{d^2}{dt^2} \left(\frac{f^2_0}{R_0^2}\right) + 3H\frac{d}{dt} \left(\frac{f^2_0}{R_0^2}\right) \right] = \\ \nonumber && - \frac{3}{2}\frac{4n^2+13n+10}{4n^2+5n+1}R_0.
\end{eqnarray}
As expected, the D'Alembertian terms cannot be neglected, as they are comparable to $R_0 \sim H^2$.

Several values for the exponent $n$ have been evaluated in previous studies, ranging from studies of hydrostatic equilibrium \cite{solarBP} or spherical collapse \cite{collapse} to galactic \cite{dm1BP} and cluster \cite{dm2BFP} dark matter, dark energy \cite{BFP} and post-inflationary preheating \cite{reheating}. All scenarios assumed a linear $f^1(R) = 2\kappa R$, except for the latter --- where $f^1 (R) = 2\kappa (R + R^2/6M^2) $ (the so-called Starobinsky inflation).

In all of these studies, it has been argued that any particular power-law form for the NMC represents the dominant behaviour of a more evolved function $f^2(R)$ in each regime ({\it i.e.} typical scalar curvature associated with the context under scrutiny, from astrophysics to  cosmology). As an example, a particular set $(n,R_n)$ that accounts for {\it e.g.} galactic dark matter was shown to be irrelevant  to implement a generalized preheating after inflation (and {\it vice-versa}). This argument is also used concerning the plethora of forms used for the curvature term in $f(R)$ theories.

The same reasoning should apply here: for completeness, the full set of power-laws contributions considered in the mentioned studies should be used, that is, $f^2(R) = \sum_i \left( \frac{R}{R_i} \right)^{-i}$. However, since this quantity (and its derivatives) must be evaluated at its cosmological value $R = R_0(t)$, it suffices to retain the cosmologically dominant term, as studied in Ref. \cite{BFP}. Thus, the results here obtained cannot be used to constraint the power-law NMC functions used to account for astrophysical scenarios (including galactic and cluster dark matter).

With the above in mind, we recall the two examples presented numerically in Ref. \cite{BFP}, were
\begin{eqnarray}\label{casestudyvalues}
n=4&:&~~ t_4= \frac{t_0}{4} \rightarrow f^2_0 = \left( \frac{12}{85} \right)^4 \approx 4 \times 10^{-4}, \\ \nonumber n=10&:&~~ t_{10}= \frac{t_0}{2} \rightarrow f^2_0 = \left( \frac{3}{451}\right)^{10} \approx 10^{-22},
\end{eqnarray}
confirming that the NMC is indeed perturbative, as indicated above.

\subsection{Long range regime, $|mr| \ll 1$}

Using Eqs. (\ref{case-study}), (\ref{dalemb-1}) and (\ref{dalemb-2}), we are now able to compute the mass parameter
given by Eq. (\ref{mass-formula}), obtaining
\begin{eqnarray}\label{mass-formula-case-study}
m^2 &=& \frac{ \mu\rho^{\rm cos}+ \nu\rho^{\rm s}}{\rho^{\rm cos}+\rho^{\rm s}} R_0, \\ \nonumber \mu&\equiv& - \frac{8n^3+4n^2-18n+1}{6n(n+1)(4n+1)} , \\ \nonumber \nu&\equiv&\frac{28n^2+111n+89}{6(n+1)(4n+1)}.
\end{eqnarray}
Notice that the roots of the denominator of both $\mu$ and $\nu$ are non-positive, while the NMC used in a cosmological setting assumes a positive exponent $n$ \cite{BFP}.

\begin{figure}
 \includegraphics[width= \linewidth]{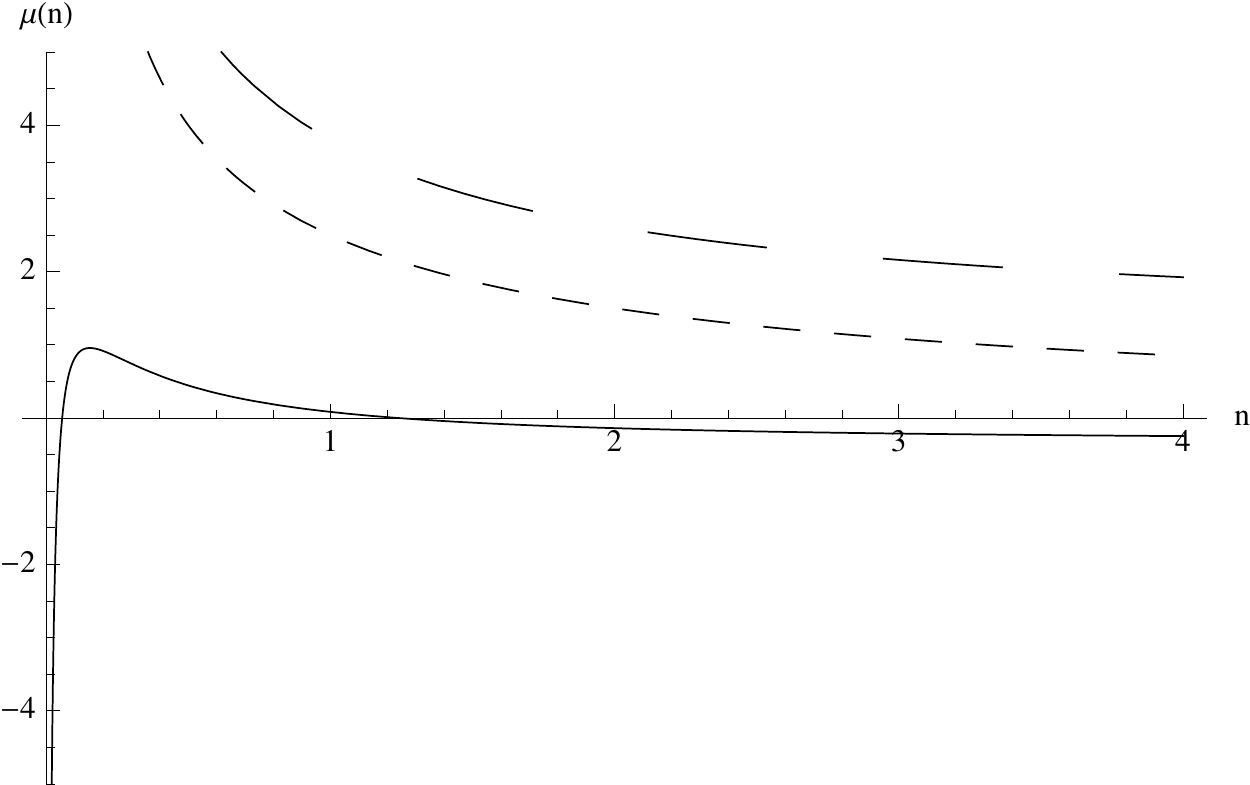}%
 \caption{Quantities $\mu(n)$ (full), $\nu(n)$ (long dash) and $\epsilon(n)$ (short dash) , defined in Eqs. (\ref{mass-formula-case-study}) and (\ref{perturbative-condition-outside-case-study}) as a function of the exponent $n$.}
\label{graphfunctions}
\end{figure}

Fig. \ref{graphfunctions} shows the variation of $\mu(n)$ and $\nu(n)$: for $n > 0$, we see that both functions are $O(10)$ or below: since $\rho^{\rm s}\gg \rho^{\rm cos}$ inside the spherical body --- except for a vanishingly thin surface layer ---, the mass parameter is given inside it by $m^2 \sim \nu R_0$ (for all values of $n$, since $\nu$ has no roots); in the outside, we have $m^2 = \mu R_0$.

For $n \sim 0$, the function $\mu$ grows to large (negative) values; if $ \mu\rho^{\rm cos} \gg \nu\rho^{\rm s}$, the mass parameter inside the spherical body is given by
\begin{equation}
m^2 \approx \frac{\rho^{\rm cos} }{\rho^{\rm s}} \mu R_0.
\end{equation}
Since $\mu \sim - 1/6n$ for $n \sim 0$, the validity of the long-range regime yields
\begin{equation}
|m r | \leq |m \radiustar | \ll 1 \rightarrow n \gg \frac{\rho^{\rm cos} }{\rho^{\rm s}} \frac{ \radiustar^2 R_0}{6} \sim 10^{-66} ,
\end{equation}
using $\rho^{\rm cos} \sim 10^{-27}~{\rm kg/m^3}$, $\rho^{\rm s} \lesssim \rho^{\rm s}_0\sim 10^5~{\rm kg/m^3}$ (the central density of the Sun), $R_0 \sim (10^{14}~{AU})^{-2}$ and $\radiustar = 1.4 \times 10^9 ~{\rm m} \sim 5 \times 10^{-3}~{\rm AU} $.

By the same token, away from the spherical body we get
\begin{equation}
|m r | \lesssim |m {\rm R}_{S}|  \ll  1 \rightarrow n \gg \frac{ {\rm R}_{S}^2 R_0}{6} \sim 10^{-25},
\end{equation}
a stronger constraint than the one above, but extremely mild nonetheless.

\subsection{Newtonian regime}

The previously discussed Eq. (\ref{Newt-limit}) provides the condition for the validity of the Newtonian regime adopted in this study. Using the previous expressions Eq. (\ref{case-study}) and (\ref{rhocosmo}), we find that $f^2_{R0}\rho^{\rm cos}(t)/\kappa = {\rm const.}$, and this condition can be recast as
\begin{eqnarray}\label{Newt-limit-case_study}
 \nonumber && \left\vert \frac{\kappa}{f^2_{R0}\rho^{\rm cos}(t)} - 1 \right\vert \rho^{\rm cos} = \left( 3 + \frac{1}{2n} \right) \rho^{\rm cos}(t) \gg \rho^{\rm s}(r) \rightarrow \\ && n \ll \frac{\rho^{\rm cos}}{2\rho^{\rm s}} \sim 10^{-33},
\end{eqnarray}
which is incompatible with the constraint $n \gg 10^{-25}$ required for the long-range condition $|mr| \ll 1$ to be valid outside the spherical body. Nevertheless, we can not yet conclude that the Newtonian limit is not valid
for $n \gg 10^{-25}$, since we have still to check the validity of Eq. (\ref{condlinear}), {\it i.e.} our assumptions that terms nonlinear in $R_1$ are negligible in the Taylor expansions of $f^i(R)$ and $f^i_R(R)$. This will be the subject of Section D.

\subsection{PPN parameter $\gamma$}

If nonlinear terms in $R_1$ were negligible in the Taylor expansions Eqs. (\ref{condlinear}), then the result
of the preceding section implies that the Newtonian approximation would not be valid in the Solar System,
whenever $|mr| \ll 1$, {\it i.e.} $n \gg 10^{-25}$.
Thus we cannot rely on the result presented here for its impact at Solar System scales, {\it i.e.} the expression for the PPN $\gamma$ parameter, Eq. (\ref{gamma}).

\subsection{Perturbative regime, $R_1 \ll R_0$}

We now check our assumption that $R_1 \ll R_0$. At the end of Section \ref{sec:sol-Psi}, in order to check
such an assumption outside the spherical body, we have used the inequality
$GM_{\rm S} \slash \radiustar \ll 1$, where $G$ is the effective gravitational constant defined
in Eq. (\ref{G-constant}).
However, the result of subsection B shows that in the long-range regime $|mr| \ll 1$ we can not rely on
the validity of Newtonian limit, so that we are prevented from using the effective gravitational constant $G$
in this way.
Hence, in order to estimate the ratio $R_1 \slash R_0$, in the sequel we resort to Newton's gravitational constant
$G_N$, which we recall is defined by $\kappa = c^4/16\pi G_N$.

\subsubsection{Outer solution}

We first assess the validity of the perturbative condition $R_1 \ll R_0$ outside the spherical body.

From Eq. (\ref{etadef}), we find that a perturbative coupling $f^2_0 \ll 1$ yields $\eta(t) \approx -1/3$. Using Eqs. (\ref{R1-solution}), (\ref{rhocosmo}) and (\ref{constantdeceleration}), we get
\begin{eqnarray}\label{perturbative-condition-outside-case-study}
\frac{R_1}{R_0} &=& -\frac{R_0}{24\pi n(n+1)f^2_0\rho^{\rm cos}} \frac{M_{\rm S} }{ r} = \\ \nonumber && -\frac{1+4n}{18\pi n f^2_0t_0^2 \rho^{\rm cos}} \frac{M_{\rm S} }{ r} = -\epsilon \frac{G_NM_{\rm S} }{ 3 r}, \\ \nonumber && \epsilon \equiv  \frac{1+4n}{n (1+n)}.
\end{eqnarray}
We see that the function $\epsilon$, plotted in Fig. \ref{graphfunctions}, has no positive roots, but diverges at $n=0$. Thus, we get
\begin{eqnarray}
R_1 \ll R_0 \rightarrow n \gg \frac{G_NM_{\rm S} }{ 3 \radiustar} \approx 7.1 \times 10^{-7},
\end{eqnarray}
a stronger constraint that those obtained in the preceding section.

\subsubsection{Inner solution}

We now assess the validity of the perturbative condition $R_1 \ll R_0$ inside the spherical body.

We address Eq. (\ref{perturbativecondition}): using Eq. (\ref{U-equation-simplified-inside-dim-solution})
and $\eta \approx -1/3$, the former reads
\begin{equation}\label{perturbative-condition-inside-case-study}
\frac{R_1}{R_0} = \frac{\rho^{\rm s}}{\rho^{\rm cos} + \rho^{\rm s}} \frac{ 1 - z(t) w(x)}{1+n} ,
\end{equation}
defining the dimensionless form function
\begin{equation}
w(x) \equiv \frac{\rho^{\rm s}_0}{\rho^{\rm s}}\sum_{i=0} \frac{a_i}{i+2} \left( 1- \frac{x^{i+2}}{i+3} \right),
\end{equation}
and coupling
\begin{eqnarray}\label{defz}
z(t) &\equiv& \frac{\eta(t) \radiustar^2}{2 f^2_{R0} } \approx \\ \nonumber && \frac{1}{6n}  \left(\frac{4(1+4n) (1+n) }{3} \right)^{n+1} \left(\frac{\radiustar}{t}\right)^2 \left(\frac{t_n}{t}\right)^{2n} = \\ \nonumber && \frac{1+4n}{n} \frac{4\pi}{3}G_N \radiustar^2\rho^{\rm cos}(t) ,
\end{eqnarray}
again using $\eta \approx -1/3$ and Eq. (\ref{constantdeceleration}). The above may be recast as
\begin{equation}
z(t) = \frac{1+4n}{n} \frac{4\pi}{3} \frac{G_NM_{\rm S}}{\radiustar} \frac{\rho^{\rm cos} \radiustar^3}{M_{\rm S}} ,
\end{equation}
clearly showing that, since $G_NM_{\rm S}/\radiustar \sim 2\times 10^{-6}$ and $\rho^{\rm cos} \radiustar^3 /M_{\rm S} \sim 10^{-31}$, $z(t)$ is vanishingly small unless $n \ll 10^{-37}$.

At the surface of the spherical body, $x=1$, we have $\rho^{\rm s} = \rho^{\rm s}_0 \sum_{i=0}a_i=0$, so that
\begin{eqnarray}
\label{perturbative-condition-boundary-case-study}
\frac{R_1}{R_0} &=& -\frac{ z(t)\rho^{\rm s}_0}{ (n+1) \rho^{\rm cos} } \sum_{i=0} \left( \frac{a_i}{i+3} \right) = \\ \nonumber && - \frac{1}{4\pi} \frac{z(t)}{ (n+1) \rho^{\rm cos} } \frac{M_{\rm S}}{\radiustar^3} ,
\end{eqnarray}
and, using Eq. (\ref{defz}), Eq. (\ref{perturbative-condition-outside-case-study}) is matched at the surface, as expected.

\begin{figure}
\includegraphics[width= \linewidth]{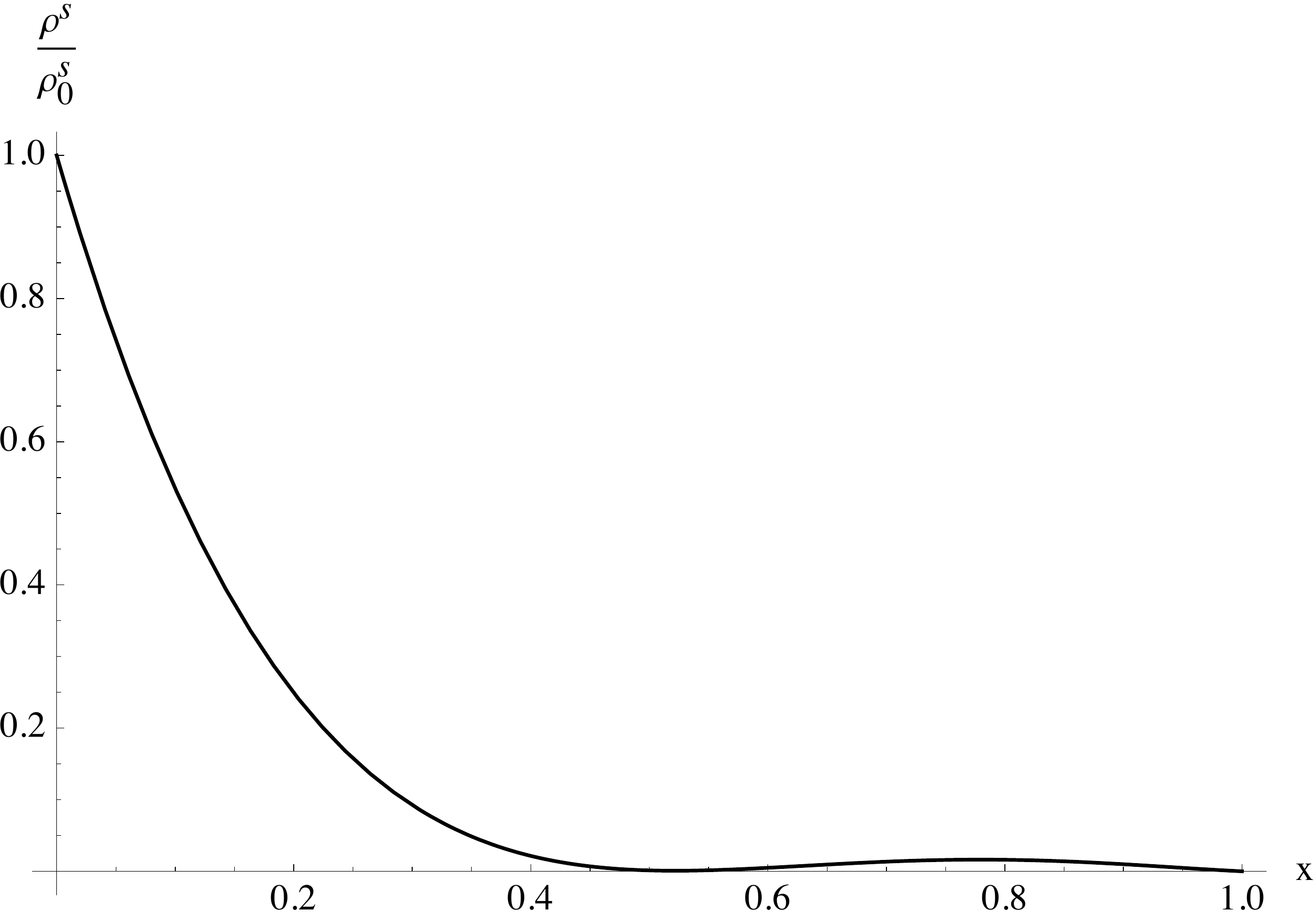}%
\caption{Fourth-order approximation of the density profile inside the spherical body (Eq. (\ref{NASAprofile}).}
\label{NASAdensity}
\end{figure}

To assess the behaviour inside the spherical body, we consider the following model of the density profile of the Sun \cite{NASAprofile}
\begin{equation}\label{NASAprofile}
\rho^{\rm s}(r) = \rho^{\rm s}_0 \left( 1 - 5.74 x+ 11.9x^2 - 10.5x^3 + 3.34x^4\right) ,
 \end{equation}
depicted in Fig. \ref{NASAdensity}. As discussed below, the overall result is not qualitatively affected by the specific density model. Notice that this fourth-order expression obeys the constraint $\rho^{\rm s}(\radiustar)=0$ and $(d\rho^{\rm s}/dr)(\radiustar) \simeq 0$.

\begin{figure}
\includegraphics[width= \linewidth]{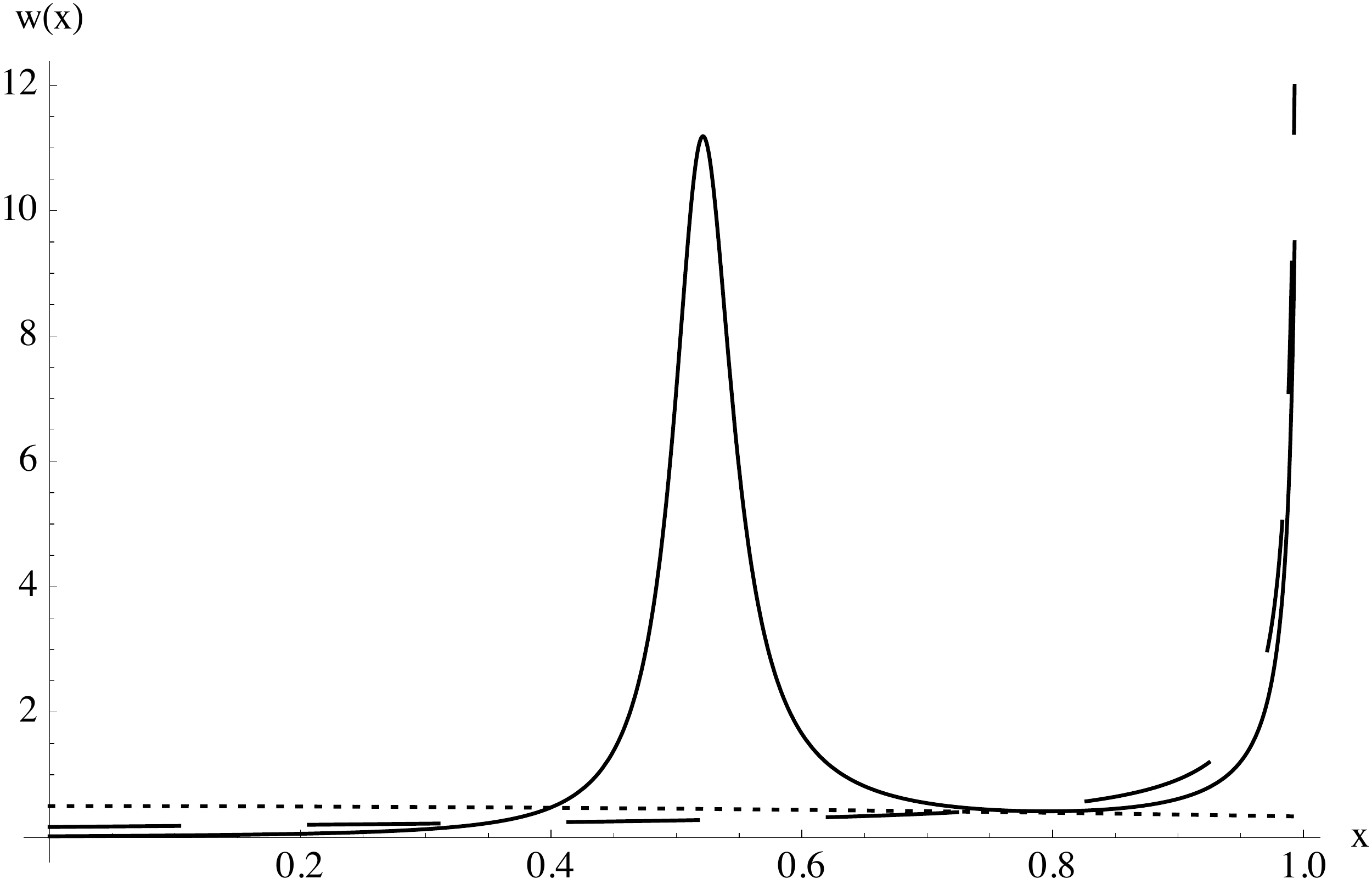}%
\caption{Form functions $w(x)$ for the density profile Eq. (\ref{NASAprofile}) ($w_4(x)$, full), linear ($w_1(x)$, dashed) and constant profile ($w_0(x)$, dotted).}
\label{figw}
\end{figure}

The density $\rho^{\rm s}(r)$ rises beyond $\rho^{\rm cos}$ immediately after the surface of the spherical body: for the chosen density profile Eq. (\ref{NASAprofile}), we find numerically that $\rho^{\rm s} \gg \rho^{\rm cos} \rightarrow x < 1- 10^{-31}$. Thus, this thin surface layer may be safely disregarded, and Eq. (\ref{perturbative-condition-inside-case-study}) is approximated by
\begin{equation}\label{perturbative-condition-inside-case-study2}
\frac{R_1}{R_0} \approx \frac{ 1 - z(t) w(x)}{1+n} .
\end{equation}
Fig. \ref{figw} plots the form function $w(x)$ for the density profile above. For comparison, two unrealistic cases are also depicted: $w_1(x)$, obtained from a linear density $\rho^{\rm s}=\rho^{\rm s}_0 (1-x)$, and $w_0(x)$, derived from a constant one, $\rho^{\rm s}=\rho^{\rm s}_0$.

Clearly, the peak around $x \sim 0.5$ for the form function $w_4(x)$ derived from Eq. (\ref{NASAprofile}) appears because the density has a minimum at $x \approx 0.52$ (which is an unphysical artifact of the fourth-order approximation). We also see that $w_1(x)$ is an approximate envelope of $w_4(x)$, {\it i.e.} presents an approximate behaviour without the aforementioned peak.

If $n \ll 10^{-37}$, a large coupling $z(t) \gg 1$ arises and we get $R_1 \approx -z(t) w(x) R_0$. This result breaks the perturbative condition underlying this work, moreover in this case the long-range condition $|mr| \ll 1$ is not satisfied.

The converse case $n > 10^{-37}$ (which comprises $n=4$ or $n=10$, the two scenarios studied in Ref. \cite{BFP}) leads to
\begin{equation}\label{perturbative-condition-inside-case-study3}
\frac{R_1}{R_0} \approx \frac{1}{1+n},
\end{equation}
which is valid for the full interior of the spherical body, with the exception of a very thin surface layer signaling the transition to the outer solution. Notice that this result is not dependent on the adopted density model, as the vanishingly small value of $z(t)$ absorbs any peaks that may arise in the form factor $w(x)$.

Eq. (\ref{perturbative-condition-inside-case-study3}) implies that the condition $R_1 \ll R_0$ is not satisfied
when $n \sim 1$ or $n < 1$.
For $n=4$, the value of the curvature perturbation $R_1$ is one fifth of the cosmological background curvature $R_0$, while $n=10$ yields a smaller $1/11$ factor. At first sight, this result allows us to validate the perturbative condition $R_1 \ll R_0$, or at least it leads to the conclusion that $n \gg 1$ --- in order to get a larger separation between $R_1$ and $R_0$. However, this is not the case: indeed, if we expand the power-law NMC Eq. (\ref{case-study}) up to third order and insert Eq. (\ref{perturbative-condition-inside-case-study3}),
\begin{eqnarray}
f^2(R) &\simeq& f_0^2\bigg[ 1 - n \frac{ R_1}{R_0} + \frac{n(n+1)}{2} \left(\frac{R_1}{R_0}\right)^2 \\ \nonumber && - \frac{1}{6}n(n+1)(n+2) \left(\frac{R_1}{R_0}\right)^3 \bigg] = \\ \nonumber &&  f_0^2\bigg[ 1 - \frac{n }{n+1} + \frac{n}{2(n+1)}  - \frac{n(n+2) }{6(n+1)^2} \bigg] ,
\end{eqnarray}
we conclude that, for any exponent $n > 1$, the non-linear terms in the Taylor expansion of $f^2(R)$ cannot be disregarded. An analogous result holds for the Taylor expansion of the function $f^2_R(R)$.
It follows that conditions Eqs. (\ref{condlinear}) are not respected.

A third possibility remains: that the coupling $z(t)$ is such that it enables a small numerator in Eq. (\ref{perturbative-condition-inside-case-study2}): this implies that $z(t) w(x) \sim 1$, which requires an approximately constant form function $w(x) \approx {\rm const.}$. However, since $z(t)$ is determined by the choice of the cosmologically relevant NMC, this would lead to an unphysical fine tuning of the form function $w(x)$, and is thus deemed unfeasible.

Given the above discussion, we conclude that the perturbative regime is not compatible with the scenario posited in Ref. \cite{BFP}, and thus the method here developed cannot be applied to constrain the latter using Solar System observables.

\subsection{Post-inflationary reheating model}

Following Ref. \cite{reheating}, we now consider the model
\begin{equation}\label{nmc-reh}
f^1(R) = 2\kappa \left( R + \frac{R^2}{6M^2} \right), \qquad f^2(R) = 2\xi\frac{R}{M^2},
\end{equation}
which adds a non-minimal coupling to the standard preheating scenario in the context of Starobinsky inflation \cite{Starob}. In Eq. (\ref{nmc-reh}), $M$ has dimensions of mass and $\xi$ is a dimensionless parameter.
The mass parameter $m^2$, defined in Eq. (\ref{mass-formula}), is proportional to $M^2$. Since $M^2$
is large in Starobinsky gravity, the condition $mr \ll 1$ is not satisfied inside the Solar System and we cannot use the present analysis to constrain the NMC model (\ref{nmc-reh}).

\section{Conclusions and Outlook}

We have analyzed the constraints that the NMC Eq. (\ref{case-study}) should fulfill in order to be consistent with the regimes considered in this work. This is summarized as follows:
\begin{itemize}
\item Long-range regime $|m| r \ll 1$ within the Solar System, leading to $n \gg 10^{-25}$;
\item Newtonian approximation, leading to $n \ll 10^{-33}$;
\item Perturbative regime $R_1 \ll R_0$, only viable if $z(t) w(x) \sim 1$ (see Eq. (\ref{perturbative-condition-inside-case-study2})), thus leading to an unphysical fine tuning of the density profile inside the spherical body.
\end{itemize}

The lack of validity of the perturbative regime leads us to conclude that the mechanism proposed in Ref. \cite{BFP} cannot be constrained or excluded by the method developed in the present paper.

This result, however, is not specific to the Sun or similar objects, but is characteristic of any relevant spherical body of astrophysical scale for which the weak field approximation can be used.

Nevertheless, this study provides a relevant set of tools with which to assess the local impact of proposals for a perturbative power-law NMC driving the accelerated expansion of the Universe. Notice that the procedure can also be applied for a NMC that does not follow a power-law form, as long as its temporal variation (and of its derivatives) is of the order of $H^2 $.

Of course, in what concerns the cosmological context, a new set of issues associated with the treatment of cosmological perturbations must be considered in order to address the impact of the NMC (see Ref. [10]).

\appendix
\section*{Appendix}

We compute here the solution of Eqs. (\ref{eqpsi}). We set
\begin{eqnarray}
R_1(r,t) &=& A(t) \frac{M}{r} \quad r\geq \radiustar, \\ \nonumber
A(t) &=& \frac{1 +  f^2_0 -2 f^2_{R0}R_0 - 6 \square f^2_{R0}}
{12\pi \left( f^1_{RR0} - 2 f^2_{RR0}\rho^{\rm cos} \right)}.
\end{eqnarray}
Using the divergence theorem, for $r\geq \radiustar$ we have
\begin{equation}\label{Psi-derivative}
2r^2 \frac{d\Psi_1}{dr} =\frac{1}{2}A(t)M\left(\radiustar^2 - r^2\right) - \int_0^{\radiustar} R_1(r,t)r^2dr .
\end{equation}
From the definition of function $U$ and the generalized mean value theorem for integrals we have
\begin{eqnarray}
&& \int_0^{\radiustar} R_1(r,t)r^2dr = \\ \nonumber && \frac{1}{f^1_{RR0} - 2 f^2_{RR0}\left( \rho^{\rm cos} + \rho^{\rm s}(\xi)\right)}
\int_0^{\radiustar} U(r,t)r^2dr,
\end{eqnarray}
for $\xi\in(0,\radiustar)$.

Using Eq. (\ref{U-equation-simplified}) and the divergence theorem, for $r\leq \radiustar$ we have
\begin{eqnarray}
\frac{dU}{dr} &=& \eta(t)\frac{m(r)}{4\pi r^2} + 2 f^2_{R0}\frac{d\rho^{\rm s}}{dr},\\ \nonumber
m(r) &=& \int_{B_r}\rho^{\rm s}(x)d^3x,
\end{eqnarray}
where $B_r$ is the ball of radius $r$ centered at the center of the star. Imposing the condition
$\lim_{r \to 0}U(t,r)r^3 =0$, repeated integration by parts yields
\begin{eqnarray}
&& \int_0^{\radiustar} U(r,t)r^2dr = \\ \nonumber && \frac{1}{3}U(t,\radiustar)\radiustar^3 - \frac{\eta(t)}{12\pi}\int_0^{\radiustar}m(r)rdr +
\frac{1}{2\pi} f^2_{R0}M.
\end{eqnarray}
Substituting the previous results into Eq. (\ref{Psi-derivative})  yields, for $r\geq \radiustar$:
\begin{eqnarray}
\Psi_1(r,t) &=& -\frac{1}{24\pi r}\,\frac{\eta(t)}{f^1_{RR0} - 2 f^2_{RR0}\left( \rho^{\rm cos} + \rho^{\rm s}(\xi)\right)} \\ \nonumber &&\times
\left( M \radiustar^2 + \int_0^{ \radiustar}m(r)rdr + \frac{1}{4\pi} f^2_{R0}M \right) \\ \nonumber && - \frac{1}{4}A(t)M\left( \frac{\radiustar^2}{r} + r \right) + C_1,
\end{eqnarray}
where $C_1$ is an integration constant. Now we estimate the various contributions to $\Psi_1(r,t)$. We have
\begin{eqnarray}
&& I = \\ \nonumber &&  \frac{\vert\eta(t)\vert}{24\pi r}\,\frac{1}
{\left\vert f^1_{RR0} - 2 f^2_{RR0}\left( \rho^{\rm cos} + \rho^{\rm s}(\xi)\right) \right\vert}
\int_0^{\radiustar}m(r)rdr \leq \\ \nonumber
&& \frac{\vert\eta(t)\vert}{48\pi}\,\frac{1}
{\left\vert f^1_{RR0} - 2 f^2_{RR0}\left( \rho^{\rm cos} + \rho^{\rm s}(\xi)\right) \right\vert}
\,\frac{M \radiustar^2}{r},
\end{eqnarray}
from which, using $r\geq \radiustar$, it follows
\begin{equation}
I \leq \frac{r^2}{8}\left\vert \frac{f^1_{R0} - 2 f^2_{R0}\rho^{\rm cos}}
{f^1_{RR0} - 2 f^2_{RR0}\left( \rho^{\rm cos} + \rho^{\rm s}(\xi)\right) }\right\vert
\,\left\vert \frac{\eta(t)}{\omega(t)} \right\vert \,\left\vert \Psi_0 - C_0 \right\vert,
\end{equation}
where $\omega(t)$ has been defined in Eq. (\ref{G-constant}). Using Eq. (\ref{rhocos-small}), we have
\begin{equation}\label{approx-0}
\left\vert 2 f^2_{R0} \right\vert \left(\rho^{\rm cos}(t) + \rho^{\rm s}(r)\right) \ll
\left\vert f^1_{R0} \right\vert, \qquad r \leq \radiustar.
\end{equation}
Thus, the following approximation can be used:
\begin{eqnarray}\label{approx}
&& \left\vert f^1_{R0} - 2 f^2_{R0}\rho^{\rm cos}(t) \right\vert \simeq \\ \nonumber &&
\left\vert f^1_{R0} - 2 f^2_{R0}\left( \rho^{\rm cos}(t) + \rho^{\rm s}(\xi) \right) \right\vert,
\end{eqnarray}
from which, using condition (\ref{constraint}), it follows that
\begin{eqnarray}
I  &\lesssim &\frac{r^2}{8} \left\vert \frac{f^1_{R0} + 2 f^2_{R0}\LL_m(\xi,t)}
{f^1_{RR0} + 2 f^2_{RR0}\LL_m(\xi,t)} \right\vert
\,\left\vert \frac{\eta(t)}{\omega(t)} \right\vert \,\left\vert \Psi_0 - C_0 \right\vert
 \\ \nonumber &\ll& \left\vert \Psi_0(r,t) - C_0 \right\vert,
\end{eqnarray}
where we have used $\vert\eta(t)\slash\omega(t)\vert \sim 1$ and
$\left\vert R_0 \right\vert r^2 \sim H^2 r^2 \ll 1$, for the current Hubble parameter and for $r$ of the order of
Solar System scales. Analogously, we have
\begin{eqnarray}
II &=& \frac{\vert\eta(t)\vert}
{\left\vert f^1_{RR0} - 2 f^2_{RR0}\left( \rho^{\rm cos} + \rho^{\rm s}(\xi)\right) \right\vert}
\, \frac{M \radiustar^2}{24\pi r} \\ \nonumber &\ll& \left\vert \Psi_0(r,t) - C_0 \right\vert.
\end{eqnarray}
By the same token, we get
\begin{eqnarray}
III &=&  \vert A(t) \vert \frac{M \radiustar^2}{4r} \simeq \\ \nonumber &&
\frac{\vert\delta(t)\vert}{\left\vert f^1_{RR0} - 2 f^2_{RR0}\rho^{\rm cos} \right\vert}
\,\frac{M \radiustar^2}{48\pi r} \ll \\ \nonumber && \left\vert \Psi_0(r,t) - C_0 \right\vert,
\end{eqnarray}
where we have neglected $\square f^2_{R0}$, and we have set
$\delta(t) = 1 +  f^2_0 -2 f^2_{R0}R_0$. We can then estimate
\begin{eqnarray}
IV &=& \frac{1}{4} \vert A(t) \vert Mr \simeq \\ \nonumber &&
\frac{r^2}{8}\left\vert \frac{f^1_{R0} - 2 f^2_{R0}\rho^{\rm cos}}
{f^1_{RR0} - 2 f^2_{RR0}\rho^{\rm cos}}\right\vert \left\vert\frac{\delta(t)}{\omega(t)}\right\vert
\left\vert \Psi_0(r,t) - C_0 \right\vert,
\end{eqnarray}
so that the estimate $IV \ll \left\vert \Psi_0(r,t) - C_0 \right\vert$ follows in the same way as for the term $I$.

It remains to consider the term
\begin{equation}
V = \frac{\left\vert  f^2_{R0} \right\vert}{4\pi} \,
\frac{1}{\left\vert f^1_{RR0} - 2 f^2_{RR0}\left( \rho^{\rm cos} + \rho^{\rm s}(\xi)\right) \right\vert}
\, \frac{M}{r}.
\end{equation}
Using Eq. (\ref{Newt-limit}) and integrating over the volume of the star, we have
\begin{equation}
\left\vert  f^2_{R0} \right\vert M \ll \frac{4\pi}{3}
\left\vert f^1_{R0} - 2 f^2_{R0}\rho^{\rm cos}(t) \right\vert \radiustar^3,
\end{equation}
from which, using Eq. (\ref{approx-0}) and condition (\ref{constraint}), we have, for $r \geq \radiustar$:
\begin{eqnarray}
V &\ll& \frac{1}{3}\left\vert \frac{f^1_{R0} - 2 f^2_{R0}\left( \rho^{\rm cos} + \rho^{\rm s}(\xi) \right)}
{f^1_{RR0} - 2 f^2_{RR0}\left( \rho^{\rm cos} + \rho^{\rm s}(\xi)\right) }\right\vert \,r^2
= \\ \nonumber &&\frac{1}{3}\left\vert \frac{f^1_{R0} + 2 f^2_{R0} \LL_m(\xi,t)}
{f^1_{RR0} + 2 f^2_{RR0}\LL_m(\xi,t)}\right\vert \,r^2
\sim \\ \nonumber &&\frac{1}{3}\left\vert R_0 \right\vert r^2 \sim \frac{1}{3} H^2 r^2.
\end{eqnarray}
Since the quantity $\Psi_0(r,t) - C_0$ turns out to be the Newtonian potential (see Eqs. (\ref{Psi-solution}) and
(\ref{G-constant})), we have $\vert V \vert \ll \left\vert \Psi_0(r,t) - C_0 \right\vert$
for $r$ of order of Solar System scales. Eventually we have
\begin{equation}
\Psi_1(r,t) - C_1 = I + II + III + IV + V,
\end{equation}
and, collecting the above estimates, we find that
\begin{equation}
\left\vert \Psi_1(r,t) - C_1 \right\vert \ll \left\vert \Psi_0(r,t) - C_0 \right\vert.
\end{equation}

\begin{acknowledgments}

The work of O.B. and J.P. is partially supported by FCT (Funda\c{c}\~ao para a Ci\^encia e a Tecnologia, Portugal) under the project PTDC/FIS/111362/2009.
The work of R.M. is partially supported by INFN (Istituto Nazionale di Fisica Nucleare, Italy), as part of the MoonLIGHT-2 experiment in the framework of the research activities of the Commissione Scientifica Nazionale n. 2 (CSN2).

\end{acknowledgments}



\begin{thebibliography}{99}

\bibitem{CSE}
T. Chiba, T.L. Smith and A. L. Erickcek,
Phys. Rev. D {\bf 75}, 124014 (2007).
%
\bibitem{status}
O. Bertolami and J. P\'aramos, ``The experimental status of Special and General Relativity'', to appear in {\it Handbook of Spacetime}, Springer, Berlin (2013); arXiv:1212.2177 [gr-qc].
%
\bibitem{DeT}
A. De Felice and S. Tsujikawa,
Living Rev. Rel. {\bf 13}, 3 (2010).
%
\bibitem{BBHL}
O. Bertolami, C. G. B\"{o}hmer, T. Harko and F. S. N. Lobo,
Phys. Rev. D {\bf 75}, 104016 (2007).
%
\bibitem{solarBP}
O.~Bertolami and J.~P\'aramos, Phys.\ Rev.\ D {\bf 77}, 084018 (2008).
%
\bibitem{BS}
O. Bertolami and M. C. Sequeira, Phys. Rev. D {\bf 79}, 104010 (2009).
%
\bibitem{equivalenceBP}
O. Bertolami and J. P\'aramos, {\it Class. Quant. Grav.} {\bf 25}, 245017 (2008).

\bibitem{dm1BP}
O.~Bertolami and J.~P\'aramos, JCAP {\bf 03}, 009 (2010).
%
\bibitem{dm2BFP}
O.~Bertolami, P.~Fraz\~ao and J.~P\'aramos, Phys. Rev. D {\bf 86}, 044034 (2012).
%
\bibitem{pertBFP}
O. Bertolami, P. Fraz\~ ao and J. P\'aramos, {\it JCAP} {\bf 05}, 029 (2013).
%
\bibitem{constantBP}
O.~Bertolami and J.~P\'aramos, Phys. Rev. D {\bf 84}, 064022 (2011).
%
\bibitem{reheating}
O. Bertolami, P. Fraz\~ao and J. P\'aramos,
Phys.\ Rev.\ D {\bf 83}, 044010 (2011)
%
\bibitem{BFP}
O. Bertolami, P. Fraz\~ ao and J. P\'aramos,
Phys. Rev. D {\bf 81}, 104046 (2010).
%
\bibitem{BLP}
O. Bertolami, F. S. N. Lobo and J. P\'aramos,
Phys. Rev. D {\bf 78}, 064036 (2008).
%
\bibitem{collapse}
J. P\'aramos and C. Bastos,
Phys. Rev. D {\bf 86}, 103007 (2012).
%
\bibitem{limitBM}
O. Bertolami and A. Martins, Phys.\ Rev.\ D {\bf 85}, 024012 (2012).
%
\bibitem{timelikeBF}
O. Bertolami and R. Z. Ferreira, Phys.Rev. D {\bf 85}, 104050 (2012).
%
\bibitem{CDTT}
S.M. Carroll, V. Duvvuri, M. Trodden and M.S. Turner,
Phys. Rev. D {\bf 70}, 043528 (2004).
%
\bibitem{HMV}
K. Henttunen, T. Multam\" aki and I. Vilja,
Phys. Rev. D {\bf 77}, 024040 (2008).
%
\bibitem{ChibaReview}
T. Chiba,
Prog. Theor. Phys.  {\bf 126}, 993 (2011).

\bibitem{WMAP9}
C. L. Bennett {\it et al.}, arXiv:1212.5225 [astro-ph.CO].
%
\bibitem{NASAprofile}
{\it http://spacemath.gsfc.nasa.gov}.
%
\bibitem{Starob}
A.A. Starobinsky,
Phys. Lett. B {\bf 91}, 99 (1980).
%
\end{thebibliography}
\end{document}